\documentstyle[preprint,revtex]{aps}
\math-with-secnums
\begin{document}
\hskip -2\wd0\copy1
\begin{title}
Twisted Boundary Conditions  and the Adiabatic
 Ground State for the Attractive XXZ Luttinger Liquid.
\end{title}
\vskip 1truecm
\author{Naichang Yu and Michael Fowler}
\vskip 1truecm
\begin{instit}
                          Department of Physics
                          University of Virginia
                          Charlottesville
                          VA 22901
\end{instit}
\centerline{May 18 1992}
\vskip 1truecm
\begin{abstract}
The one-dimensional attractive lattice fermion gas equivalent to the
Heisenberg-Ising spin 1/2 chain is studied for a ring geometry
threaded by magnetic flux. We find that for charged fermions having
interaction strength $\Delta=\cos(\pi / p)$ with $p$ {\it
noninteger}, the adiabatic ground state  is periodic in the
magnetic flux threading the chain, with period 2 flux quanta, as
found by Shastry and Sutherland for the repulsive case. We find
that, at particular values of the threading field, a sequence of
initially zero-energy bound states form at the Fermi surface during
the adiabatic process, the largest containing $[p]$ (the
integer part of $p$) fermions. This largest bound state moves
around to the
other Fermi point and sequentially unbinds. We find Berry's Phase
for the whole process  to be $[p] \pi$.  For $p$
{\it integer}, as $\Phi$ increases, eventually all the particles in
the system go into bound states of size $[p]$. The period in this
case is of order the size of the system.
The charge carrying mass of the fermions is calculated by finding
the
energy of the adiabatic ground state with twisted boundary
conditions.
\end{abstract}
\newpage
\section{Introduction}
     There has been some interest recently in examining one
dimensional integrable systems with twisted boundary conditions.
These conditions are defined, for a system on a ring, by requiring
that on taking any one of the particles completely around the ring,
the wavefunction does not match up, but is off by a constant phase
factor $e^{i\Phi}$ (so the usual periodic boundary conditions
correspond to $\Phi =0$). A model of interacting fermions with
these boundary conditions may be a reasonable model for mesoscopic
rings with only one, or perhaps a few, electron channels, threaded
by a magnetic flux, and some of its features may be related to
properties of edge states in quantum Hall droplets.  For integrable
systems, the twisted boundary
conditions have one very appealing feature---for any given value of
$\Phi$, the equations remain exactly solvable by the Bethe ansatz
method. This is therefore one of the few perturbations one can
impose on an integrable system that does not spoil the
integrability, so its effect on the states of the system can be
computed very precisely.  For example, it is easy to compute the
ground state energy accurately as a function of the twist angle,
and, indeed, to follow the development of the state in detail.
Recently, Shastry and
Sutherland\cite{suth901,suth902} (SS) examined
the fermion gas formulation of the Heisenberg-Ising spin one-half
chain (XXZ model) with twisted boundary conditions, in the
repulsive regime.  The XXZ spin chain, first systematically
analyzed by Yang and Yang\cite{yang66}, was also the model with
which Haldane\cite{haldane80} introduced the concept of a Luttinger
liquid.  For this system, twisted boundary
conditions
 are physically equivalent to a magnetic flux of
$\Phi/2\pi$ flux units threading the ring on which the fermions
move. SS computed the dependence of the ground
state energy on the twist angle $\Phi$ for the repulsive case, and
found the relevant term
to be
$D\Phi^{2}$, where $D$ is essentially the inverse of the
charge-carrying mass in the system.  The physical point here is
that
varying the twisting angle slowly in time is equivalent to varying
the magnetic flux, which generates an electric field around the
ring, so the energy change computed is the response of the system
to
this slowly varying electric field.  Therefore the low frequency
conductivity of the gas, which is determined by $D$, can be
extracted from a Bethe ansatz calculation of the ground state
energy with the twisted boundary conditions. SS also uncovered some
very interesting behavior in the adiabatic ground state wave
function as a function of twist angle $\Phi$.  In particular, they
found extremely rapid variation of the highest particle's bare
momentum as
$\Phi$ varied near $2\pi$.

  In the present paper, we focus mainly on the {\it attractive}
regime
for the XXZ fermion gas.   We begin our analysis of these systems
in section two by
numerically diagonalizing some small spin one-half Heisenberg-Ising
chains with various interaction strengths, and finding the full
energy spectrum as a function of the twisting angle. We  find  that
the
adiabatic period of the ground state is determined by when the
evolving state is first {\it nondegenerate} for $\Phi$ an integer
multiple of $2\pi$.

     In section three we briefly review the rather strange
behavior of the Bethe ansatz eigenvalues as functions of the twist
angle $\Phi$ found by Shastry and
Sutherland, and connect it with other properties of Bethe ansatz
systems.  We then go on to the attractive fermion gas, where we
find
some novel analytic behavior in the motion of the
eigenvalues as $\Phi$ varies.  For example, we find that as $\Phi$
passes through particular values, which are independent of the size
of the system, successively larger zero-energy bound states
form at the right fermi point, up to a maximum of $[p]$ fermions.
This maximum bound state then moves very rapidly over to the left
fermi point, where successive unbinding occurs
as $\Phi$ increases towards $4\pi$, at which value the system
returns to the ground state, for the general case.
We show the rapid motion of the largest bound state is closely
related to
the perfect shielding of string
excitations of certain sizes in Bethe ansatz systems.  It is also
interesting to note that the sequence of fermi-point zero-energy
bound states represent the quantum group structure of the $XXZ$
model described by Pasquier and Saleur\cite{pasquier} .  We find
that
SS's novel result, that the ground state of the
repulsive fermion gas is periodic in $\Phi$ with period $4\pi$, is
also true in general for the attractive gas
{\it except} at those values of the attractive potential at which
a new breather-type excitation is generated, in which case more
complicated
behavior arises.  The coefficient
$D$ giving the conductivity is easily extended into the attractive
regime.

     Recently, Korepin and Wu\cite{korepin} calculated the change
in phase of the
ground state wave function (the Berry\cite{berry} phase) for the
repulsive gas
on increasing the
boundary angle from zero to $4\pi$.  They found the
phase change to be $\pi$. In section six, we extend this result
into the attractive regime, and find that the
phase change alternates between zero and $\pi$ as the coupling
strength increases.

\section{Numerical Diagonalization of Small Systems}

To illuminate the differences between free fermion systems and
those having repulsive or attractive interactions, we have carried
out explicit diagonalizations of small systems numerically. We were
able to find the full energy spectrum with twisted boundary
conditions, and thus the adiabatic ground state  as a function of
twist angle $\Phi$.
The hamiltonian  of
the system is:\cite{yang66,des}
\begin{equation}
H = -
1/2\sum_{j=1}^{N}\left(\sigma_{j}^{x}\sigma_{j+1}^{x}+\sigma_{j}^
{y}\sigma_{j+1}^{y}+\Delta\sigma_{j}^{z}\sigma_{j+1}^{z}\right) .
\end{equation}
     The standard notation is to write $\Delta=-\cos\mu$.  The
Hamiltonian also represents spinless lattice fermions
through the Jordan-Wigner transformation,
\begin{equation}
H=-\sum_{j}(c_{j}^{+}c_{j+1}+c_{j+1}^{+}c_{j}) - 2\Delta\sum_{j}
(c_{j}^{+}c_{j}-1/2)(c_{j+1}^{+}c_{j+1}-1/2) .
\end{equation}
The twist in the boundary condition is:
\begin{equation}
\sigma_{N+1}^z = \sigma_{1}^z, \qquad \qquad  \sigma_{N+1}^x \pm
i\sigma_{N+1}^y = e^{i \Phi} ( \sigma_{1}^x \pm i\sigma_{1}^ y ) .
\end{equation}
 The spectrum clearly satisfies the relation:
\begin{equation}
E(\Phi) = E(-\Phi) = E(2\pi+ \Phi),
\end{equation}
so we need only consider the range $0 \le \Phi \le \pi$. Fig.~1a
shows  the lowest 10 adiabatic energy levels of a chain of 6 sites
and 3 fermions, with repulsive interaction strength $\Delta =-
\cos(\pi/3)$. It is easy to follow the adiabatic evolution of the
ground state. The state crosses with the first excited state (which
is also evolving) at
$\Phi = \pi$. At $\Phi=2 \pi$, the ground state has evolved into
the first excited state of the original $\Phi=0$ system. This state
is nondegenerate, and as $\Phi$
increases through a further $2\pi$, it evolves back to the ground
state. This picture is correct for any repulsive interaction, and
is the adiabatic process described by  Sutherland and
Shastry.\cite{suth902}

Let us now consider what happens for free fermions (Fig.~1b). This
is of course a trivial exercise in momentum space, where in the
ground state the fermions half-fill a Brillouin zone, and as $\Phi$
increases from zero they move steadily to the right, by one quantum
state for each $2\pi$ twist of $\Phi$, successively exiting at the
right-hand edge of the zone $k=\pi$ and simultaneously reappearing
at the left-hand point $k=-\pi$. For a system of $N$ spins it is
easy to see that the period is $2N\pi$.

By comparing Fig.~1a with Fig.~1b, we can see why the periods are
so different for free fermions and interacting fermions.  The
essential point is the degeneracy or nondegeneracy of the state the
ground state evolves to when $\Phi$ increases to $2\pi$. Since the
energy spectrum is symmetrical as a function of $\Phi$ about
$\Phi=0\pmod{2\pi}$, a nondegenerate state at such a value of
$\Phi$
must necessarily have zero slope as a function of $\Phi$, so if the
ground state evolves up to a nondegenerate state as $\Phi$
increases to $2\pi$, it must by symmetry retrace its steps
(reflected in $\Phi=2\pi$)  as $\Phi$ goes from $2\pi$ to $4\pi$,
giving period $4\pi$.  On the other hand, if, as for the
noninteracting case, the symmetry requirement about $\Phi=2\pi$ is
satisfied by having two states degenerate there having opposite
slopes as functions of $\Phi$, following one of these adiabatically
means that the energy continues to increase as $\Phi$ passes
through $2\pi$, and any turnaround must take place later.

We now turn to chains with attractive interactions between
fermions. Fig.~1c shows the lowest 6 energy levels for the same
chain with attractive interaction $\Delta =  \cos( 2 \pi / 5)$,
 $(\mu=0.6\pi)$. We
see that between $0 \le \Phi \le 2\pi$, the ground state crosses
with 3 energy levels (the 2nd and 3rd excited states are
degenerate), and at $\Phi=2\pi$ the ground state has evolved into
the 4th excited state.
However, this state is
again nondegenerate, and consequently the state begins to evolve
back to the
ground state which it reaches at $\Phi=4\pi$. Thus the adiabatic
period is
again $4\pi$. The evolution of the ground state for any  chain
having
$\Delta =  \cos( \pi /p)$,  $2<p<3 $, is similar to this case.

 However, as shown in Fig.~1d,  for a chain with
$\Delta=\cos(\pi/3)$, that is to say, $\mu=2\pi/3$,
the state that the ground state evolves into at $\Phi=2\pi$ is
again {\it degenerate}  with another state, and the ground state
will evolve to higher excited states before it evolves  back. The
adiabatic period  in this case will be proportional to $N$, the
length of the chain, as it was for the noninteracting case.

The spectrum for general attraction with $\Delta=\cos(\pi/p)$ is
similar to one of the above two cases. For $p$ noninteger, the
adiabatic period is $4\pi$, while for $p$ integer, the period is
proportional to the length of the chain.

Numerical diagonalization of small systems yields not only the
spectrum, but also the actual wavefunctions as functions of twist
angle $\Phi$. Thus it is possible to compute Berry's phase for the
cyclic process, the (nondynamic) phase gained by the wavefunction
during the adiabatic period.  It is necessary to establish a
reference set of wavefunctions which are periodic with period
$4\pi$,
\begin{equation}
\Psi (\Phi+4 \pi) = \Psi( \Phi) ,
\label{4pi}
\end{equation}
then Berry's phase can be computed using the formula:
\begin{equation}
\gamma =  {\rm Re}  \left[ i \int_{0}^{4\pi} d\Phi
{{<\Psi(\Phi)|{\partial \over {\partial \Phi} } |\Psi (\Phi)> }
\over {<\Psi(\Phi) |\Psi (\Phi)>}} \right] .
\label{berryform}
\end{equation}

We find that provided the chains are sufficiently long that there
are more fermions than the number $[p]$ (the integer part of $p$)
then Berry's phase is $\pi$ if $[p]$ is an odd integer, and $0$ if
$[p]$ is even.

\section{Bethe Ansatz Analysis}
In the standard Bethe-ansatz approach, modified for the twisted
boundary conditions, any eigenfunction of the Hamiltonian is a
superposition of plane waves\cite{korepin,korbook}:
\begin{equation} |\Psi> = \sum_{x_1=1}^N \cdots  \sum_{x_N=1}^N
\chi ( x_1,\ldots, x_M) \prod_{j=1}^M \sigma_{x_j}^- | \uparrow> ,
\label{wavefunction}
\end{equation}
where $| \uparrow>$ is the state with all spins up, and $\sigma
_x^-$ is the spin
lowering operator at site $x$. The function $\chi$ has the form:
\begin{eqnarray}
\chi (x_1,\ldots,x_M| p_1,\ldots, p_M) = \left( \prod_{M \ge b
 > a \ge 1} \right ) \epsilon (x_b-x_a)  \nonumber\\
\sum_Q (-1)^Q \exp \left[ i\sum_{a=1}^M {x_a p_{Q_a} } \right
]
\exp \left[ {i/2} \sum_{M \ge b > a \ge 1} \theta (p_{Q_b} ,p_
{Q_a} ) \epsilon (x_b-x_a) \right] ,
\label{chi}
\end{eqnarray}
where $\epsilon (x) $ is the sign function, the summation is
with respect to permutations $Q$ of the momenta $p$, and $\theta$
is
the two particle
phase shift, which for two particles with momenta $p$, $q$ is
\begin{equation}
\theta(p,q) = 2\arctan\left(\frac{\Delta\sin((p-
q)/2)}{\cos((p+q)/2) - \Delta\cos((p-
q)/2)}\right)
\end{equation}

  The wavefunction is determined by a set of $M$ quantum numbers
${I_j}$, each corresponding to the total phase change (kinetic plus
phase shifting on passing other
particles)  for one of the particles going completely around the
system, so for the twisted
boundary condition,
\begin{equation}
N p_j + \sum_{k=1}^M \theta(p_j,p_k) = 2 \pi I_j + \Phi .
\label{baeq}
\end{equation}
The standard approach is to parameterize the momenta in terms of
rapidities $\lambda$ by
\begin{equation}
p(\lambda) = -i \ln \left(-{ {\sinh\frac{1}{2}( \lambda-i\mu  ) }
\over {\sinh\frac{1}{2}
 (\lambda+ i \mu ) } }\right) ,
\label{p(lambda)}
\end{equation}
because the interparticle phase shift depends only on the rapidity
difference, making the equations easier to handle.  Note that with
the rapidity variable defined as above, and the usual convention
for the branches of the logarithm, zero momentum corresponds to
zero rapidity.  As the momentum increases from zero to $\pi-\mu$,
the rapidity goes from zero to infinity (monotonically).  Further
increase in the momentum causes the rapidity variable to jump to
the $i\pi$ line, whereupon its real part monotonically decreases to
zero, reaching $\lambda=i\pi$ at momentum $p=\pi$.

It is worth noting at this point that the energy of a single spin
wave excitation above the ground state $|\uparrow \rangle $ is
given by
\begin{equation}
E(k) = 2(\Delta - \cos k).
\end{equation}
Here $\Delta=-\cos\mu$, so the single particle excitation has
negative energy if its pseudomomentum lies in the range $|k|<\pi-
\mu$, which is just the range for which the rapidity $\lambda$ lies
on the real axis.

Transforming to the rapidity variables, the expression $\theta
(p,q)$ for the particle-particle phase shift becomes:
\begin{equation}
\theta(\lambda_{1},\lambda_{2})=i \ln \left(-{ {\sinh\frac{1}{2}(
\lambda_{1}-\lambda_{2}-2i\mu  ) }
\over {\sinh\frac{1}{2}
 (\lambda_{1}-\lambda_{2}+ 2i \mu ) } }\right),
\label{theta(lambda)}
\end{equation}
a function of the difference only, and obviously very similar to
the expression for the momentum above, but with opposite sign. The
phase shift as written
here is zero for $\lambda_{1}=\lambda_{2}$, and $\pi-2\mu$ if
$\lambda_{1}$ is infinite, $\lambda_{2}$ finite.

In the rapidity variables,
the Bethe-ansatz equations (3.4)
for allowed momenta with the twisted boundary condition are
equivalent to:
\begin{equation}
\left(- {{\sinh\frac{1}{2}(\lambda_i +i\mu)}
          \over { \sinh\frac{1}{2}(\lambda_i-i\mu) }} \right) ^N
   = e^{-i\Phi}\prod _{j\not= i} \left(-
          {{ \sinh\frac{1}{2}(\lambda_i-\lambda_j+2i\mu) }
          \over { \sinh\frac{1}{2}(\lambda_i-\lambda_j-2i\mu) }}
\right) ,
\end{equation}
where $\Phi$ is the twisting angle.

Let us briefly review some
properties of the ground state and lowlying states of the system
for the standard
periodic boundary condition $\Phi=0$.
The roots of the Bethe ansatz equations are usually  grouped into
strings  which differ only in imaginary part by $2i(\pi-\mu)$ (in
the thermodynamic limit) and center on the real axis or the $i\pi$
line:
\begin{equation}
\lambda_{i,k}^n = \lambda_i^n + i(\pi-\mu)( n+1-2k)+i\pi(
 1- \nu _n ) + \delta_k \qquad k=1,\ldots,n .
\end{equation}
     Here $\nu_n = 0,1$ is called the parity of the string, and
$\delta_k$ is the     deviation---for a finite system---from the
evenly-spaced string found in the infinite limit.  This deviation
is
of order $1/N$ at most. Such a group of $n$ roots is called an
$n^+$
($n^-$) string if it is centered on the real axis($i\pi$ line),
so the sign denotes the value of the string parity $\nu_n$.

The ground state for $\Phi=0$ for the {\it noninteracting} case is
just the sea of filled negative-energy single-particle states, that
is, the band is half-full, from $-\pi/2$ to $\pi/2$.  The
appropriate set of quantum numbers in equation (\ref{baeq}) is,
trivially in this case,
\begin{equation}
 I_j = -(M-1)/2,-(M-3)/2,\ldots,(M-1)/2.
\end{equation}
It turns out that this same set of integers also gives the ground
state energy for the interacting case.  The total energy is still
given by summing over the single particle energies $2(\Delta-\cos
p_j)$ (although the allowed values of the $p_j$'s of course shift
from the free values), so the lowest energy state has as many
$p_j$'s as can be put in the negative energy range $|p_j|<\pi-\mu$.
We shall see shortly that this number is always $N/2$, as would be
expected from the antiferromagnetic nature of the system, since it
corresponds to the total spin in the $z$-direction being zero.

The elementary excitations above this ground state are given by
either holes in the fermi sea, or strings centered on the
$i\pi$-line. For these excitations to have normalizable (bare)
wavefunctions in
the thermodynamic limit, the string must fit into a strip of width
$2\pi i$ (ignoring certain exceptional long strings discovered by
Korepin\cite{korepin79} which are not relevant to the analysis
below). Since the
spacing between members of a string is $2i(\pi-\mu)$, the allowed
string lengths for a given value of the anisotropy parameter $\mu$
are:
\begin{equation}
n = 1,2,\ldots \left[\frac{\pi}{\pi-\mu}\right] +1
\end{equation}
It is convenient to introduce a variable $p$ defined by:
\begin{equation}
p = \frac{\pi}{\pi - \mu}.
\label{p}
\end{equation}
Denoting the integer part of $p$ by $[p]$ in the standard way, the
total number of allowed string lengths is $[p]+1$.  However, not
all these string lengths correspond to the usual type of excitation
expected in such systems.  Specifically, the two longest strings in
the series induce a backflow in the sea of filled negative energy
states which precisely cancels their bare energy and momentum in
the thermodynamic limit.  For a string with this peculiar property,
there is only phase space available for a single string above the
ground state, even for a large system.  This is essentially because
of the quasifermionic properties of Bethe-ansatz excitations, and
the fact that all such strings have the same total momentum to
order $1/N$.  These properties are well-illustrated by numerical
analysis of systems with twisted boundary conditions: very small
changes of the twist angle, which move ordinary excitations through
distances of order $1/N$ in momentum space, will drive one of these
strings across the whole Brillouin zone.  For the repulsive fermion
gas considered by Shastry and Sutherland, the $1^{-}$ and $2^{-}$
strings have this property, the $1^{-}$ string being a single root
on the $i\pi$-line.  For the noninteracting gas, there are no such
strings because there is no phase shifting and hence no backflow.
For the weakly attractive gas, the $2^{-}$ and $3^{-}$ strings are
of this type, but the $1^{-}$ string is an ordinary excitation,
corresponding to the breather in the sine-Gordon
system\cite{korepin79,thacker}.  As $\mu$
increases from $\pi /2$ (noninteracting gas) towards $\pi$, the
number of allowed string lengths increases, but for any $\mu$ the
two longest strings
are always of the peculiar type, the rest being breather-like
physical excitations.
\section{adiabatic ground state}
We use the equation
\begin{equation}
N p_j + \sum_{k=1}^M \theta(p_j,p_k) = 2 \pi I_j + \Phi .
\end {equation}
together with the appropriate set of quantum numbers
\begin{equation}
 I_j = -(M-1)/2,-(M-3)/2,\ldots,(M-1)/2.
\end{equation}
to determine how the momenta in the adiabatic ground state vary as
the angle $\Phi$ increases from zero.

We have solved these equations numerically for reasonably large
systems, and we present our results in this section together with
some analytic interpretation.

Let us first  look at the repulsive regime $-1<\Delta<0$ considered
by SS. The $\Phi=0$ ground state has the $p_i$ symmetrically
arranged in the interval $-(\pi-\mu), (\pi-\mu)$ which maps on to
the whole real axis in the $\lambda$-plane.  It turns out to be
very helpful to picture what is happening in both the momentum and
the rapidity variables simultaneously.

In fig 2, we plot $p$ as a function of $\lambda$ for $\lambda$ both
on the real axis and on the $i\pi$ line, using
equation (\ref{p(lambda)}).  On the real axis, $p$ is
zero for $\lambda$ zero, and odd in $\lambda$. It increases
monotonically as a function of $\lambda$ from $-(\pi-\mu)$ at
$\lambda=-\infty$ to $(\pi-\mu)$ at $\lambda=+\infty$, most of the
increase occurring in a region of order $\mu$ near the origin.
Further increase in $p$ corresponds to $\lambda$ moving back along
the $i\pi$-line from $i\pi+\infty$, so for $p=\pi$, $\lambda=i\pi$,
and
for $\lambda=i\pi-\infty$, $p=2\pi-(\pi-\mu)$, $p$ having therefore
picked up
$2\pi$ by going around this closed circuit, since the path
encircles a logarithmic branch cut as a function of
$\lambda$.

The phase shift $\theta$ between two excitations, given as a
function of
the relative rapidity by equation (\ref{theta(lambda)}), is a very
similar curve, but with {\em opposite}
sign.  It decreases from $(\pi-2\mu)$ at $\lambda=-\infty$ to $-
(\pi-2\mu)$ at $\lambda=+\infty$, then further decreases to $2\pi-
(\pi-\mu)$ as $\lambda$ moves along the $i\pi$ line from
$i\pi+\infty$ to $i\pi-\infty$.

Let us now consider how the rightmost root, $p_M$, which is very
close to $(\pi-\mu)$ (just below it) at $\Phi=0$, moves as $\Phi$
increases from zero. It satisfies the equation:
\begin{equation}
N p_M + \sum_{k=1}^{M-1} \theta(p_M,p_k) = 2 \pi\left(\frac{M-
1}{2}\right) + \Phi
\end {equation}
and as $\Phi$ increases, $p_M$ is pushed up towards $(\pi-\mu)$.
Glancing at the rapidity space representation, it is clear that
this root gets far away from the others, since $(\pi-\mu)$
corresponds to infinite $\lambda$, so that the relative phase
shifting with the other roots tends to the value $-(\pi-2\mu)$, and
the value of $\Phi$ at
which $p_M=\pi-\mu$ is given by, using $N=2M$,
\begin{equation}
2M(\pi-\mu)-(M-1)(\pi-2\mu)=2\pi\left(\frac{M-1}{2}\right)+\Phi
\end{equation}
that is, $p_M=\pi-\mu$ when $\Phi=2(\pi-\mu)$.

Note that this result is {\em independent} of the size of the
system.

{}From the momentum space point of view, it is clear that further
increasing $\Phi$ will increase $p_M$ further, or, switching back
to rapidity space again, $\lambda_M$ will jump to the $i\pi$ line
and begin to move to the left.  The subsequent motion is easier to
understand in rapidity space because of the much simpler form of
the phase shift.

We see from the graph of $p$ as a function of $\lambda$ that for
$\lambda_M$ to go from $i\pi+\infty$ to $i\pi-\infty$ implies an
increase in $p_M$ of $2\mu$.  However, as $\Phi$ is twisted to
bring this about, we note that each time $\lambda_M$ passes over
another root $\lambda$ on the real axis, there is a relative phase
shifting of $4\mu$ (almost all taking place when $\lambda_M$ is
within a distance
of order $\pi/2-\mu$ of the point directly over $\lambda$).  From
the equation above for $p_M$, this
phase shifting causes $p_M$ to increase by an amount
$(1/2M).(4\mu)$
above that directly attributable to the increase in $\Phi$.  Thus,
as $\lambda_M$ moves from $i\pi+\infty$ to $i\pi-\infty$, it gets
a total momentum boost from passing over all the other roots of
$(M-1).(1/2M).(4\mu) \simeq 2\mu$. But this is essentially all the
momentum increase needed for $\lambda_M$ to go from $i\pi+\infty$
to $i\pi-\infty$, as stated above. It follows that only a very
slight
increase in $\Phi$ is needed to drive $\lambda_M$ over all the
other roots.  Since from symmetry $\lambda_M$ is at $i\pi$ for
$\Phi=2\pi$, the motion past all the other roots must take place
very close to this value.  Another interesting point is that as
$\lambda_M$ is boosted forward by passing over a root $\lambda_i$,
the $\lambda_i$ is moved in the opposite direction by an equal
amount, essentially equal to the local root spacing on the real
axis.  The effect of this is to cause the motion of $\lambda_M$
along the $i\pi$ line to be accompanied by a ripple of compression
(equal to one extra root, spread over a length of order $\pi/2-
\mu$) in the roots on the real axis.

Since less than a $2\pi$ twist in $\Phi$ sends the root the entire
length of the $i\pi$ line, there is no phase space available to put
more such excitations on this line, provided the fermi sea is fully
occupied, to give the necessary boosts as described above.  This is
a finite system formulation of the well-known fact that in the
large system limit, such excitations have exactly zero energy and
momentum, and confirms that for the ground state of the system,
there is no room for more than $N/2$ roots.

For the noninteracting case, $\mu=\pi/2$ , there is of course no
phase shifting between the particles, so as $\Phi$ twists, all the
$p_i$'s move together in a train, returning to the ground state at
$\Phi=2N\pi$.

Let us now move to the attractive case.  The behavior is very
different because for $\mu > \pi/2$ the phase shift function
({\ref{theta(lambda)})
changes sign, it is now an {\em increasing} function of relative
rapidity, going on the real axis from $-(2\mu-\pi)$ at $\lambda=-
\infty$ to $(2\mu-\pi)$ at $\lambda=\infty$, then continuing to
increase for $\lambda$ moving back on the $i\pi$ line from
$i\pi+\infty$.

On initially increasing $\Phi$ from zero, $\lambda_M$ reaches
$\infty$ at $\Phi=2(\pi-\mu)$, just as for the repulsive case (and
indeed for the noninteracting case).  As before, it then jumps to
the $i\pi$ line and begins to move back.  But there is a dramatic
change of behavior as it comes within range of the next highest
root $\lambda_{M-1}$, which is still
on the real axis.  The change in the sign of
the phase shift means that the presence of $\lambda_{M-1}$ {\em
slows
down} the leftward progress of $\lambda_M$ along the $i\pi$ line.
In fact, for $\lambda_M$ to draw level with $\lambda_{M-1}$ would
require a phase shift of $\pi-(2\mu-\pi)$, that is, $\Phi$ would
have to increase from $2(\pi-\mu)$ to $4(\pi-\mu)$.  This is
exactly what happens, but at the same time $\lambda_{M-1}$ is
boosted forward by the increasing $\Phi$ and the phase shifting
from $\lambda_M$, so it drives $\lambda_M$ back and they come
together at infinity for $\Phi = 4(\pi-\mu)$.  It is trivial to
check that this gives an exact solution of the equations for $p_M,
p_{M-1}$, bearing in mind that these two roots have a mutual phase
shifting of $\pm \pi$ when one is directly above the other.

What happens as $\Phi$ now increases through $4(\pi-\mu)$ is most
easily visualized in $p$-space. The two largest roots, $\lambda_M$
and $\lambda_{M-1}$, approach $p=(\pi-\mu)$ from opposite sides,
moving asymptotically at the same rate, meeting there when
$\Phi=4(\pi-
\mu)$, then separating in opposite directions parallel to the
imaginary axis, forming a bound state.  This is just the behavior
of roots of a quadratic equation as a coefficient is varied through
the point where they coincide, and is not difficult to verify
analytically.  This rather natural behavior looks odd in the
rapidity plane.  The two roots, having (asymptotically) the same
real part
and imaginary parts 0 and $i\pi$, go to infinity, where they jump
together discontinuously by
$i\pi/2$ in the imaginary direction, and then begin to move
together back from infinity as a two-string centered on the $i\pi$
line.

To follow the further development of the system once the two-string
has formed, it is simplest to back up slightly and consider the
behavior of two-strings over the whole range of $\mu$ discussed so
far. The basic two-string has two roots centered at a point on the
$i\pi$ line, having pure imaginary separation $2i(\pi-\mu)$. (This
spacing can vary somewhat for finite systems, especially for a
two-string near infinity.  This does not affect our conclusions
below).  For small $\mu$, the two roots are actually close to the
real axis ($\mu$ on either side).  The momentum of the two-string,
$p_{2^-}(\lambda)$, is given by just adding the momenta of the two
roots from equation(\ref{p(lambda)}):
\begin{equation}
p_{2^-}(\lambda) = -i \ln \left(-{ {\sinh\frac{1}{2}( \lambda-2i\mu
) }
\over {\sinh\frac{1}{2}
 (\lambda+ 2i \mu ) } }\right) ,
\end {equation}
The variable $\lambda$ appearing in the above equation is the point
on the real axis above which the string is located.  (This defines
the string uniquely---there cannot be both $2^+$ and $2^-$ strings
in the same system.) In the repulsive regime $\mu<\pi/2$, the
(bare) momentum  $p_{2^-}$ of the two-string increases
monotonically as a function of $\lambda$ from $-(\pi-2\mu)$ at
$\lambda=-\infty$ to $(\pi-2\mu)$ at $\lambda=\infty$, a total
change of $2(\pi-2\mu)$. At $\mu=\pi/2$, the two-string completely
vanishes, having zero bare energy and momentum for any
$\lambda$---this is the free-fermion system.  For the attractive
regime $\mu>\pi/2$, $p_{2^-}(\lambda)$ has the same analytic form
as
in the repulsive regime, but now is a {\em decreasing} function of
$\lambda$, going from $2\mu-\pi$ at $\lambda=-\infty$ to $-(2\mu-
\pi)$ at $\lambda=+\infty$.

The phase shift of the two-string
at $\lambda$ with a real root at $\lambda_1$ is given by adding the
phase shifts from the two components, using
equation(\ref{theta(lambda)}), giving:
\begin{equation}
\theta_{2}(\lambda,\lambda_{1})=i \ln \left(-{ {\sinh\frac{1}{2}(
\lambda-\lambda_{1}-i\mu  ) }
\over {\sinh\frac{1}{2}
 (\lambda-\lambda_{1}+ i\mu ) } }.-{ {\sinh\frac{1}{2}(
\lambda-\lambda_{1}-3i\mu  ) }
\over {\sinh\frac{1}{2}
 (\lambda-\lambda_{1}+3i\mu ) } }\right)
\label{theta2}
\end {equation}

Let us consider how this phase shift varies for fixed $\lambda_1$
as $\lambda$ traverses the real axis.  First taking $\mu$ to be
small, we see that $\theta_2$ is a product of two phase functions,
both decreasing (recall $\theta$ has opposite sign to $p$), one
from $\pi-\mu$ at $\lambda=-\infty$ to $-(\pi-\mu)$ at
$\lambda=+\infty$, the other from $\pi-3\mu$ to $-(\pi-3\mu)$, to
give a total phase shift on traversing the whole line of $4\pi-
8\mu$. On increasing $\mu$ through $\pi/2$ into the attractive
regime, this value for the total phase shift remains correct until
$\mu=2\pi/3$, at which value the second phase factor in the above
expression for $\theta_2$ becomes identically zero, in a way
reminiscent of the single particle phase shift at $\mu=\pi/2$.  In
fact, the phase shift $\theta_2$ behaves very similarly---the total
phase shift from the second term switches discontinuously from
$2\pi$ to $-2\pi$ as $\mu$ passes through $2\pi/3$.

After this brief review of the properties of the momentum $p_{2^-}$
and the phase shift $\theta_2$ of a $2^-$ string, we are ready to
see what happens to the $2^-$ string which forms at infinity as
$\Phi$ increases through $4(\pi-\mu)$ for the attractive case,
$\mu>\pi/2$.

The equation for the momentum of the two-string is:
\begin{equation}
2M p_{2^-} + \sum_{k=1}^{M-2} \theta_{2}(p_{2^-},p_k) = 2
\pi\left(\frac{M-
1}{2}+\frac{M-3}{2}\right) +2\Phi
\end {equation}
The motion of this $2^-$ string in the attractive regime
$\pi/2<\mu<2\pi/3$ turns out to be very similar to that of the $1^-
$ string in the repulsive regime $0<\mu<\pi/2$ already discussed
above.  Working in the rapidity variables, as $\Phi$ increases
beyond $4(\pi-\mu)$, the $2^-$ string is driven to the left from
$\lambda=\infty$.  From the above equation, and using the fact that
going from $\lambda=+\infty$ to $\lambda=-\infty$ means a momentum
increase of $4\mu-2\pi$ for the $2^-$ string, but at the same time
the phase shifts $\theta_2$ each decrease by $8\mu-2\pi$, we see
that only a slight increase in $\Phi$ is needed to drive the two
string past all the real roots, and it is easy to verify that for
$\Phi=2\pi$ it is on the imaginary axis, symmetrically placed with
respect to the real roots.  This means that the ground state is
periodic in $\Phi$ with period $4\pi$ in this range of coupling.

This description of the motion of the roots as $\Phi$ increases is
correct provided $\mu<2\pi/3$, but if $\mu=2\pi/3$, things change
completely.  The motion for this value is most simply understood by
assuming initially that the two-strings have the canonical spacing
(that is, neglecting finite size effects) in which case the phase
shift of a two-string with a root on the real axis is given by
(\ref{theta2}).  Note that the terms involving $3i\mu$ disappear
completely for $\mu=2\pi/3$, and the remaining phase shifting from
the real roots actually slows down the leftward progress of a
two-string as $\Phi$ increases, although it does not stop it.  As
$\Phi$ increases from zero to $2\pi$, the two rightmost roots go to
infinity and form a two-string as previously discussed, which then
moves to the left by an amount of the order (in momentum) of the
spacing between the real roots.  This two-string passes the next
root at exactly $\Phi=2\pi$, at which point it has exactly the
canonical spacing, ensuring the correct phase shift factor.
We note in passing that this $\Phi=2\pi$ configuration is the
mirror image of the $\Phi=-2\pi$ configuration, so has the same
energy but is not the same state.  This is the degeneracy discussed
in section two for small systems, and only occurs when
$\mu=n\pi/(n+1)$ with $n$ integer.
Increasing $\Phi$ by a further $2\pi$, the next two real roots form
a two-string in the same way.  After an increase in $\Phi$ of
$M\pi$, all the roots are in two-strings symmetrically arranged
about the origin.
 A further increase of $M\pi$ takes the system back to the original
ground state.  Actually this analysis is oversimplified because the
two-strings have spacing that varies with position, so the $3i\mu$
terms in $\theta_2$ do not in general disappear.  However,
numerical analysis confirms that overall, the picture is still
correct.

For $\mu>2\pi/3$, the phase shift function $\theta_2$ varies from
$4(\pi-\mu)$ at $\lambda=-\infty$ to $-4(\pi-\mu)$ at
$\lambda=+\infty$.  This means that as the two-string begins to
move to the left after formation at $\Phi=4(\pi-\mu)$, it is slowed
down and actually pushed back to $\infty$ by the real root
$\lambda_{M-2}$, and at $\Phi=6(\pi-\mu)$, they come together to
form a three-string, which then moves rapidly to the left, lying on
the imaginary axis for $4\Phi=2\pi$, and again giving a system with
a ground state periodic in $\Phi$ with period $2\pi$.

The changes in the process as $\mu$ is increased further are most
conveniently discussed in terms of the variable $p=\pi/(\pi-\mu)$
introduced in (\ref{p}).
For $n<p<n+1$
with n integer,  we find that at $\Phi=2j\mu$, $j=1\ldots n$,
there are jumps of the $j$ roots with $Re(\lambda)=\infty$. Before
the jump, there are $j$ equally distanced roots between
$Im(\lambda)=\pi$ and $Im(\lambda)=-(j-1)\pi/(j+1)$. The set of $j$
roots then jump at the same time
in the imaginary direction by $-\pi/j$. Note the total change of
the
imaginary part of all the roots is always ${\pi}$. After making the
final jump at $\Phi=2n\mu$, these set of roots move to the
imaginary axis at $\Phi=2\pi$, forming an exact
$n$ string. At this point, the remaining $1^-$ strings are
positioned
symmetrically with respect to the imaginary axis.
Obviously, the period is always $4\pi$, independent of the length
of the chain.

The process for $p=n$ is different in that the roots will {\em all}
gather into
$n-1$ strings, provided we take for convenience a system of size
$N$ divisible by $2(n-1)$. The period is $8N\pi/(n-1)$,
proportional to $N$.

It is interesting to consider what happens as we approach the
ferromagnetic isotropic limit $\mu=\pi$.  For a system of size
$N=2M$, when $p>M$ we find that on twisting $\Phi$ to $2\pi$, all
the roots gather in a single $M$ string which then moves to the
imaginary axis. This configuration is of course very reminiscent of
the lowest energy state (for total $z$-spin zero) of the
ferromagnetic system approaching the isotropic point from the other
side.

\section{The Energy of the adiabatic ground state}

The $XXZ$ fermion system we are considering here was actually the
system used by Haldane\cite{haldane80} to introduce the concept of
a Luttinger liquid, although he included an external magnetic field
in the $z$-direction, equivalent to a chemical potential for the
fermions. As discussed by Haldane, the low-energy states of the
system are well described by a Hamiltonian of the form:
\begin{equation}
H=\sum\omega_{q}b^{+}_{q}b_{q}+\frac{\pi}{2N}(v_{M}(M-
M_{0})^{2}+v_{J}J^{2})
\end{equation}
where the first boson term corresponds to excitations near the
fermi points, absent in the states we are considering here.  The
other terms give the variation of system energy with the current
$J$ (an even integer, contributing momentum $k_{F}J$) and the
number of particles $M$, where $M_{0}$ is the number in the lowest
energy state for a given external field, and hence is equal to
$N/2$ for our case of an antiferromagnetic system in zero external
field.  Extending the results of Yang and Yang to the case of non-
zero current, Haldane found\cite{haldane81}:
\begin{equation}
v_{M}=2\frac{(\pi-\mu)}{\mu}\sin\mu,  \qquad
v_{J}=\frac{\pi^{2}\sin\mu}{2\mu(\pi-\mu)}
\end{equation}
It follows that these
coefficients, which have the dimensions of velocities, satisfy
$v_{M}v_{J}=v^{2}$, where $v$ is the sound velocity, so
$\omega_{q}=v|q|$ for the boson excitations above. This
relationship establishes equivalence with the Luttinger model.

These same results were derived independently by other
workers\cite{woynarovich}, as
part of a general investigation of finite size effects in Bethe
Ansatz systems, since these lead to asymptotic correlation
functions, on using conformal field theory methods. The results are
often
expressed in terms of Korepin's dressed charge\cite{bogo86}, which
is just the
square root of the velocity renormalization $(v_{J}/v)$ in
Haldane's work.

The relevance of these results to the present work is that twisting
the boundary angle $\Phi$ increases the current linearly in $\Phi$,
so
we expect that the lowest state energy as a function of $\Phi$ to
be given by replacing $J$ by $\Phi/\pi$ (since from the definition
of $J$ above, increasing $\Phi$ by $2\pi$ shifts all the quantum
numbers over by one space and so increases $J$ by 2),
\begin{equation}
E_{0}(\Phi)={\frac{\pi}{2N}} \left\{ 2\frac{(\pi-\mu)}{\mu} \sin
\mu \left( M-
\frac{N}{2}\right)^{2}+\frac{\pi^{2}\sin\mu}{2\mu(\pi-
\mu)} \left( \frac{\Phi}{\pi} \right) ^2 \right\}
\label {E_0}
\end{equation}
where the first term would be identically zero for the process we
are considering, since $M=N/2$.

Actually it is not completely obvious that this equation is valid
for $\Phi$ going from zero to $2\pi$.  The derivations mentioned
above were all based on the assumption that as the current
increases from zero, the Bethe Ansatz roots stay on the real axis.
The half filled band is the one case where that assumption is
incorrect for $\Phi$ less than $2\pi$.  For the repulsive case
($\mu<\pi/2$), SS verified (\ref{E_0}) numerically over the full
range $|\Phi| \leq 2\pi$, and mentioned that the results were
confirmed by a Wiener-Hopf analysis over that range.

For the attractive case, we have done extensive numerical
calculations and find (\ref{E_0}) to be true for $|\Phi| \leq
2\pi$.  Table I shows one of these calculations. We solved the BA
equations numerically and calculated the energy for finite systems,
then made an extrapolation to $N \rightarrow \infty$, assuming that
$\Delta E \sim 1/N$. The agreement between the numerical
calculation and (\ref{E_0}) is excellent.

In fact, it is easy to find the energy analytically for the special
twist angles $\Phi=2j(\pi-\mu)$, with $j$ integer and $2j(\pi-
\mu)<2\pi$. For this angle, $j$ of the roots are at infinity, and
their phase shifting of the remaining roots exactly compensates
that from the external twisting through $\Phi$, so the other roots
form the configuration corresponding to the ground state for
$S^{Z}=j$, or $|M-N/2|=j$ in (\ref{E_0}), and zero momentum.

We see from (\ref{E_0}) that the two states: $\Phi=2j(\pi-\mu)$,
$|M-N/2|=0$; and $\Phi=0$, $|M-N/2|=j$ have the identical
energy---they differ only in having a zero-energy bound state of
$j$
particles at the fermi point.

Finally, the energy can be found exactly for $\Phi=2\pi$.  For this
value, if $m<\pi/(\pi-\mu)<m+1$, there is an exact $m$-string on
the imaginary axis, centered on the $i\pi$ line, and the remaining
$(N/2)-m$ roots are in a symmetrical configuration on the real
axis (this $m$ is just the $[p]$ of section three). The energy of
this state can be found by the usual methods,
and again confirms (\ref{E_0}).

\section{BERRY'S PHASE}

     In this section, we extend the Berry phase result of Korepin
and Wu\cite{korepin} from  the repulsive case to the attractive
case. Their
finding was that, for the repulsive $XXZ$ system in the ground
state, on increasing the boundary condition angle $\Phi$ through
$4\pi$ as described above so that the system returns to the
original ground state, there is a net phase change (not counting
the ordinary dynamical phase) of $\pi$.

Korepin and Wu refer to the phase change in addition to the
dynamical phase  as Berry's phase although, strictly speaking, this
is a slightly different situation from that discussed by
Berry\cite{berry}, in
which the Hamiltonian moves adiabatically around a closed path in
parameter space, and every eigenstate returns to itself, with a
phase factor.  For the process we are considering, a general state
of the system will certainly not return to itself when $\Phi$ is
increased by $4\pi$, but only after an increase of $2N\pi$.

In fact, it is fairly straightforward to extend the Berry phase
result to the attractive regime once the relative motions of all
the roots have been mapped out.  The basic strategy, just as with
the numerical work on small systems in section two, is to establish
a reference set of wavefunctions having period $4\pi$ in $\Phi$,
then compute Berry's phase using \ref{berryform}.  We begin with
the ground state wave function of the form \ref{wavefunction},
\ref{chi} and carefully track all the changes in the momenta and
phase shifts as $\Phi$ increases through $4\pi$.  We find that
$\chi$ has an overall phase change resulting from this process of
zero or $\pi$.  If it is zero, the $\chi(\Phi)$ form a suitable
reference set of wavefunctions to use in \ref{berryform} for
finding the Berry phase.  If it is $\pi$, $\chi$ must be multiplied
by a phase factor $e^{i\Phi/4}$ to give the right periodicity.

We first present the calculation for the attractive
interaction $\Delta=\cos(\pi/p)$, in the range $2 \le p \le 3$. To
construct a wavefunction
continuous in $\Phi$, the crossing of the cuts of the momentum and
phase shift in the rapidity space must be considered. The cut of
the momentum
function is the line connecting the points $i \mu$ and
$i(2\pi-\mu)$ in the
complex rapidity plane. According to section IV the two moving
roots never
cross the cut, because for the backward part of their motion in
rapidity space, they are in a two-string and never get close enough
to the line ${\rm Im}\lambda=\pi$.
If we denote the initial momenta as $\{p_j^i \}$ and the final
momenta
as $\{p_j^f \}$, it follows that
\begin{equation}
p_N^f = p_1^i, \qquad p_{N-1}^f = p_2^i, \qquad p_j^f = p_{j+2}^i
{}.
\end{equation}
The cut for the phase shift is  between   $2i\mu$ and
$2i(\pi-\mu)$. It is
illuminating to consider the movement of the
difference $  \lambda_{N,N-1} = \lambda_N-\lambda_{N-1}$, in the
rapidity plane. It starts on the positive real axis,
and moves to infinity when $\lambda_N$ moves to infinity, then it
jumps to the $i\pi$ line and moves to the left, until when both
$\lambda_N$ and
$\lambda_{N-1}$ are at infinity, when  $ \lambda_{N,N-1}= i \pi$.
$\lambda_{N,N-1}$  then moves on the imaginary axis as the two
roots move as a $2^-$ string and the distance between the two
changes, until
the two roots move to minus infinity, and root $\lambda_{N-1}$
jumps down
to the real axis, $\lambda_{N, N-1}$ will then move to the left on
the $i\pi$ line. The difference clearly moves across the cut, so we
have:
$\theta( p_{N}^f, p_{N-1}^f) = \theta ( p_1^i, p_{2}^i )+2\pi$
The phase shifts between the other rapidities can be similarly
analyzed, giving
\begin{eqnarray}
\theta( p_N^f, p_j^f) = \theta ( p_1^i, p_{j+2}^i ), \nonumber\\
\theta( p_{N-1}^f, p_j^f) = \theta ( p_2^i, p_{j+2}^i ), \\
.\nonumber
\end{eqnarray}
Replacing the momentum and phase shift in (\ref{wavefunction}),
(\ref{chi}) we find that each term in the
final wavefunction corresponds to another term in the initial
wavefunction,
with a permutation of the momenta. Due to the crossing of the phase
shift cut
in one case, and the permutation in the other case, the two terms
have the
same sign, and the wavefunction satisfies:
\begin{equation}
\chi (\Phi +4\pi ) = \chi( \Phi) ,
\end{equation}
Looking at the wavefunction, it is clear that the crossing of two
momenta
across the momentum cut or the cut of the phase shift with another
root at the same $\Phi$ will
not change the wavefunction. This is indeed the case when roots not
on the
$i\pi $ line cross the momentum cuts or the phase shift cuts with
another
root on the real axis, since they always appear in pairs symmetric
with
respect to the $i\pi$ line.
The crossing of the momentum cut of a root on the  $i\pi$ line only

happens when $m$ is odd, in which case this root also cross the
phase shift cut with every
root on the real axis. The other crossing of phase shift cuts to be
considered is the phase shift cuts between two roots that make up
the $m$
string. They pass the cut when one root on the real axis meets a
$j-1$ string
at infinity to form a $j$ string. It is easy to see that at this
point the
phase shifts between this root and every root of the $j-1$ string
pass the
cuts. In total $m(m-1)/2$ cuts are passed by the phase shifts
between the
roots that make up the string.

Considering  all the crossing of the cuts and the fact that to
reverse the order of ${1,2,3 \ldots }$, $m(m-1)/2$ permutations are
needed, we have:
\begin{equation}
\chi(\Phi+4\pi) = (-1)^m \chi(\Phi)
\end{equation}
To satisfy this condition, we define $|\Psi>$ as:
\begin{equation}
|\Psi(\Phi)> = \exp( i m \Phi )|\chi (\Phi ) >
\end{equation}

 Due to the symmetry of the wavefunction with respect to  $\Phi$
and $-\Phi$ we see that:
\begin{eqnarray}
<\chi(\Phi) | \chi( \Phi)>_{\Phi} = <\chi ( \Phi) | \chi(
\Phi)>_{-\Phi}, \nonumber \\
<\chi(\Phi) | {\partial \over {\partial \Phi}} | \chi(
\Phi)>_{\Phi}   = -
<\chi(\Phi) | {\partial \over {\partial \Phi}} | \chi(
\Phi)>_{-\Phi} .  \nonumber
\end{eqnarray}
Berry's Phase is calculated as:
\begin{equation}
\gamma = \left\{ \begin{array}{ll}
                 0    & \mbox{if $m$ is even} \\
                 \pi  & \mbox{if $m$ is odd}
                 \end{array}  \right.
\end{equation}
\newpage

\section{ACKNOWLEDGEMENTS}
The authors thank H. Frahm,  V. E.  Korepin, A. Dorsey and H. B.
Thacker for numerous useful discussions.  This work was supported
by the National Science Foundation (NSF) under grant No.
DMR-8810541.

\newpage
\figure{ The lowest 10 adiabatic energy levels for a chain of 6
sites and 3 fermions with various interactions. Fig.~1a is for a
chain with repulsive interaction $\Delta = - 1/2$,  Fig.~1b shows
the levels for free fermions. Fig.~1c shows the levels for
$\Delta=\cos(\pi/2.5)$ and Fig.~1d shows the levels for
$\Delta=\cos(\pi/3)$. For general $\Delta=\cos(\pi/p)$ with $p$
noninteger, the levels are similar to these in Fig.~1c; while for
$p$ integer, they are similar to Fig.~1d.  \label{fig1}}
\newpage
\begin{table}
\caption{$\Delta E \cdot N / \sin(\mu )$ for various twisting angle
$\Phi$ for a chain of upto 80 sites with $\mu=\pi/3.5$.
Extrapolations are obtained by VBS extrapolation. Predictions are
from (\ref{E_0})}
\begin{tabular}{ccccccc}
\multicolumn{1}{c} {\rm N}
&\multicolumn{1}{c} {$\Phi=0.4\pi$} &\multicolumn{1} {c}
{$\Phi=0.7\pi$}
&\multicolumn{1} {c} {$\Phi=1.0\pi$} &\multicolumn{1} {c}
{$\Phi=1.3\pi$}
&\multicolumn{1} {c} {$\Phi=1.6\pi$} &\multicolumn{1} {c}
{$\Phi=1.8\pi$}\\
\tableline
8 &\dec 0.63334 &\dec 1.92907 &\dec 3.90330 &\dec 6.51860 &\dec
9.72165 &\dec 12.14803 \\
16&\dec 0.62008 &\dec 1.89635 &\dec 3.86157 &\dec 6.50715 &\dec
9.82013 &\dec 12.39194 \\
24&\dec 0.61767 &\dec 1.89044 &\dec 3.85433 &\dec 6.50531 &\dec
9.83792 &\dec 12.43497 \\
32&\dec 0.61683 &\dec 1.88838 &\dec 3.85175 &\dec 6.50468 &\dec
9.84411 &\dec 12.44990 \\
40&\dec 0.61644 &\dec 1.88743 &\dec 3.85056 &\dec 6.50439 &\dec
9.84696 &\dec 12.45678 \\
48&\dec 0.61623 &\dec 1.88691 &\dec 3.84991 &\dec 6.50424 &\dec
9.84852 &\dec 12.46051 \\
56&\dec 0.61610 &\dec 1.88660 &\dec 3.84952 &\dec 6.50414 &\dec
9.84945 &\dec 12.46051 \\
64&\dec 0.61602 &\dec 1.88640 &\dec 3.84927 &\dec 6.50408 &\dec
9.85006 &\dec 12.46422 \\
72&\dec 0.61596 &\dec 1.88626 &\dec 3.84910 &\dec 6.50404 &\dec
9.85047 &\dec 12.46522 \\
80&\dec 0.61592 &\dec 1.88616 &\dec 3.84845 &\dec 6.50401 &\dec
9.85077 &\dec 12.46593 \\
\tableline
VBS &\dec 0.61575 &\dec 1.88574 &\dec 3.84898 &\dec 6.50388 &\dec
9.85077 &\dec 12.46593 \\
CONJ &\dec 0.61575 &\dec 1.88574 &\dec 3.84898 &\dec 6.50388 &\dec
9.85077 &\dec 12.46593 \\
\end{tabular}
\end{table}

\newpage
\begin{references}
\bibitem[1]{suth901} B. S. Shastry and B. Sutherland, Phys.\ Rev.\
Lett.\
{\bf 65} 243 (1990)
\bibitem[2]{suth902} B. Sutherland and B. S. Shastry, Phys.\ Rev.\
Lett.\ {\bf65} 1833 (1990)
\bibitem[3]{yang66}C. N. Yang and C. P. Yang, Phys.\ Rev.\ {\bf
147}, 303 (1966); {\bf 150}, 321, 327 (1966); {\bf 151}, 258 (1966)
\bibitem[4]{haldane80} F. D. M. Haldane, Phys.\ Rev.\ Lett.\ {\bf
45} 1358 (1980), J.\ Phys.\ {\bf C14} 2585 (1981)
\bibitem[5]{pasquier} V. Pasquier and H. Saleur, Nucl.\ Phys.\
{\bf B330} 523 (1990)
\bibitem[6] {korepin} V. E. Korepin and A. C. T. Wu, Int.\ J.\
Mod.\ Phys.\ {\bf B5} 497 (1991)
\bibitem[7]{berry} M. V. Berry, Proc.\ Roy.\ Soc.\, London, {\bf
A392} 45, (1984)
\bibitem[8]{des} des Cloizeaux and Gaudin, J.\ Math.\ Phys. {\bf
7}, 1384 (1966);des Cloizeaux, J.\ Math.\ Phys.\ {\bf 7},2136
(1966)
\bibitem[9] {thacker} H. Bergknoff and H. B. Thacker, Phys.\ Rev.\
{\bf19D} 3666 (1979)
\bibitem[10]{haldane81} F. D. M. Haldane, Phys.\ Lett.\ {\bf 81A}
153 (1981)
\bibitem[11]{woynarovich} H. J. de Vega and F. Woynarovich, Nucl.\
Phys.\ {\bf B251} [FS13] 439 (1985); F. Woynarovich and H.-P.
Eckle, J.\ Phys.\ {\bf A20} L97 (1987)
\bibitem[12]{bogo86} N. M. Bogoliubov, A. G. Izergin, and V. E.
Korepin, Nucl.\ Phys.\ {\bf B275} [FS 17], 687 (1986)
\bibitem[15]{aharonov} Y. Aharonov and J. Anadan, Phys.\ Rev.\
Lett.\ {\bf 58} 1597 (1987)

\bibitem[13]{korepin79} V. E. Korepin, Theor.\ Math.\ Phys.\ {\bf
41} 953 (1979)
\bibitem[14]{korbook} V. E. Korepin, A. G. Izergin and N. M.
Bogoliubov, Quantum Inverse Scattering Method and Correlation
Functions, Cambridge, 1992
\end{references}
\end{document}